\documentclass[prd,amsfonts,twocolumn,nofootinbib,showpacs]{revtex4}
\RequirePackage{graphicx}
\usepackage{enumerate}
\pacs{98.80Cq}

\begin{document}

\title{Particle production and dissipation caused by the
Kaluza-Klein tower}

\author{Tomohiro Matsuda}
\affiliation{Laboratory of Physics, Saitama Institute of Technology,
Fukaya, Saitama 369-0293, Japan}

\begin{abstract}
Two-step dissipation is studied in supersymmetric
 models in which the field in motion couples to bulk fields in the
 higher dimensional space.
Since the Kaluza-Klein tower of the intermediate field
changes its mass-spectrum during the evolution, there could be
back-reaction from the tower.
Then the system may eventually cause significant
 dissipation of the kinetic energy if the tower is coupled to light
 fields in the thermal bath.
To see what happens in the higher dimensional theory, we consider three
 models for the scenario, which are carefully prepared.
In these models the extension is obvious but it does not disturb the
 original set-ups. 
The third model suggests that the evolution of the
 volume moduli may feel significant friction from the Kaluza-Klein tower.
\end{abstract}

\maketitle

\section{Introduction and the model}

The study of the nonequilibrium dynamics, when a background field is
time-dependent, can be motivated in many different physics of interest. 
Focusing on the cases of cosmology, one can find vast variety of
 applications in which cosmological events are studied using
nonequilibrium dynamics.
For instance in the inflation cosmology, nonequilibrium particle 
production has been studied for particle production after and during
inflation~\cite{PR, PR-matsuda, Spot-infla}, and for 
the dissipative process in warm inflation~\cite{Berera:Model-build,
 Spot-infla, Moss-Xion, Remote-warm}.  

The nonequilibrium dynamics that leads to dissipative process during inflation
can be studied assuming that the system is coupled to the thermal bath,
or alternatively assuming physical 
decay factor in the propagator, to make the local approximation valid
for the calculation. 
These studies have been developed using 
various techniques of the quantum field theory and the quantum
statistical mechanics.
A plausible scenario of dissipation has been studied in
Ref.~\cite{Berera:Model-build, Moss-Xion} for the two-step model,
using the methods of the quantum field theory when the local
approximation can be applied.
In this paper we will study the dissipation in the two-step model, when the
Kaluza-Klein tower of the intermediate field changes its spectrum when a
field moves.

Sometimes the dissipative equation of motion has been discussed 
specifically for the warm inflation scenario and thus there might be confusion
between the criticism of the specific inflation scenario and the one for
the calculation of the dissipative coefficients.
In this paper we will not assume specific inflation background during
dissipation, since the assumption of inflation requires additional
(model-dependent) conditions for the system, to keep the thermal
equilibrium a good approximation during the evolution.
Indeed, dissipative equations may play significant
role in non-inflating cosmological scenarios such as the
curvaton~\cite{Warm-curvaton} or the Affleck-Dine
baryogenesis~\cite{Affleck-D, Baryo-matsuda}.  
The purpose of this paper is to show a clear example in which the
Kaluza-Klein tower causes significant two-step dissipation in the
low-temperature limit (i.e, when the temperature of the thermal bath is
lower than the mass of the intermediate Kaluza-Klein mode.).
Model-dependent analyses are avoided in this paper.
Previous studies on string-motivated models can be found in
Ref.\cite{KK-prev}, in which the strategies are different from ours.

Let us start with the plain model. Consider the four-dimensional
supersymmetric model with the 
interaction given by the superpotential~\cite{Moss-Xion, Berera:Model-build}:
\begin{eqnarray}
W&=& W(\Phi)+g\Phi X^2 + h X Y^2,
\end{eqnarray}
where $\Phi$, $X$ and $Y$ are superfields that contains bosonic
components $\phi$, $\chi$ and $y$, respectively~\cite{Wess-Bagger}.
In this model, $X$ is the intermediate field that can decay into 
light field $Y$.
The effective mass of the intermediate field is time-dependent
because $\Phi$ is moving.
The time-dependent mass of the intermediate field causes particle
creation and the dissipation of the kinetic energy.
We are considering the effective equation of motion for the field
$\phi$, which may contain the dissipative term when the local
approximation is valid for the nonequilibrium dynamics.
The superpotential $W(\Phi)$ is not mandatory, since the potential could
be obtained from a supersymmetry breaking term that lifts a flat
direction of the supersymmetric theory.
The auxiliary field equation of the field $X$ gives
\begin{equation}
F_X^* + 2g \Phi X + hY^2=0,
\end{equation}
which leads to interactions $\sim (gh\phi) \chi y^2$ and $\sim g^2 \phi^2\chi^2$.
These terms are necessary for the later calculation of the dissipative
coefficients.

Then, as far as the local approximation is valid, the dissipative term
appears in the equation as
\begin{eqnarray}
\ddot{\phi}+(3H+{\cal \gamma})\dot{\phi}+V_\phi&=&0,
\end{eqnarray}
where the subscript of the scalar potential $V(\phi)$ denotes the derivative
with respect to the field. 
Here ${\cal \gamma}$ is the dissipative coefficient calculated in
Ref.\cite{Moss-Xion}.
In the low-T regime ($T/m_\chi\ll 1$), the coefficient is given
by~\cite{Moss-Xion, Berera:Model-build}
\begin{eqnarray}
{\cal \gamma}& \simeq & g^2 h^4 \sum_{i=R,I}
\left(\frac{g\phi}{m_{\chi_i}}\right)^4\frac{T^3}{m_{\chi_i}^2},
\end{eqnarray}
where $m_{\chi_R}$ and $m_{\chi_I}$ are the mass of the real and the
imaginary parts of the bosonic component $\chi$.
We will have $m_{\chi_R} \simeq m_{\chi_I}$ since the split
is usually negligible.
Here the interaction $\sim h m \chi y^2$ ($m\equiv g\phi$ in the above
model) is crucial for the leading-order calculation\footnote{See
Ref.\cite{Moss-Xion} for the details of the calculation.}.

The model realizes the typical two-step dissipative process and avoids
several difficulties of the warm inflation scenario.
We thus consider the supersymmetric two-step scenario as a reference
model.
Inflation background is not necessary for the process.

In this paper we will consider the higher dimensional extension of the
above two-step model.
For the effective action that describes interaction with the Kaluza-Klein tower,
we will consider the higher dimensional supersymmetry in the four-dimensional
superspace~\cite{Hamed-4DSUSY}, in which we can introduce a Gaussian
distribution for the localization in the fifth dimension~\cite{MSSM-5D}:
\begin{equation}
f_G(y)=\frac{1}{\sqrt{2\pi}l_s}e^{-\frac{y^2}{2l_s^2}},
\end{equation}
where $l_s^{-1}$ determines the width of the localization.
The localization is needed to discuss the interaction between fields on
the brane (boundary) and the fields in the bulk.
The above Gaussian function is useful since it can explain
the natural cut-off for the Kaluza-Klein tower~\cite{MSSM-5D}.
However, just for the simplicity of the calculation,
we will use $\delta(y)$ instead 
and consider a simple cut-off at $m_{KK}\simeq M_*\simeq l_s^{-1}$.
The replacement may affect numerical coefficients but does not change
the qualitative result.

\subsection{Intermediate fields in the bulk}
We consider a five-dimensional space time metric with the fifth
dimension being compactified on a $S^1/Z_2$ orbifold.
The strategy is usually considered for obtaining four-dimensional MSSM from
the flat five-dimensional action~\cite{Hamed-4DSUSY}.
For the first model, we will place the intermediate field ($X$) in the bulk while
localizing $\Phi$ and $Y$ on the boundary.

From a four-dimensional perspective, the Kaluza-Klein towers of the
intermediate field $X$ can be rearranged in the form of ${\cal N}=2$ hypermultiplets.
Choosing $Z_2$ parity of the five-dimensional fields one can break the
${\cal N}=2$ supersymmetry to ${\cal N}=1$ supersymmetry.
In ${\cal N}=1$ four-dimensional superspace, the five-dimensional
hypermultiplet consists of the four-dimensional chiral superfields 
$X(y)$ and $X^{c}(y)$, which are both labeled by $y$ (the coordinate of
the fifth dimension).
The free action is given by~\cite{Hamed-4DSUSY}
\begin{eqnarray}
S&=& \int d^5x \left\{
\left[\bar{X^{c}}X^{c}
+\bar{X}X\right]_{\theta^2\bar{\theta}^2}
+\left(X^{c}\partial_y X|_{\theta^2}
+\mathrm{h.c}\right)
\right\}.\nonumber\\
\end{eqnarray}
We will introduce the superpotential
\begin{eqnarray}
W&=&\left[g \Phi X^2 +h XY^2\right]\delta(y),
\end{eqnarray}
which involves the bulk-boundary interactions.
The auxiliary fields for the intermediate field $X$ and $X^c$ 
are computed to be:
\begin{eqnarray}
F^*_X&=&-2g\Phi X \delta (y) -hY^2\delta(y) -\partial_y X^c\nonumber\\
F^*_{X^c}&=&\partial_y X.
\end{eqnarray}
We assume the Fourier expansion on $S^1/Z_2$:
\begin{eqnarray}
\chi&=&\frac{1}{\sqrt{\pi R}}\chi^{0}+
\sqrt{\frac{2}{\pi R}}\sum_{n=1} \cos\left[\frac{n y}{R}
\right]\chi^{(n)}\nonumber\\
\chi^{c}&=&\sqrt{\frac{2}{\pi R}}\sum_{n= 1} \sin\left[\frac{n y}{R}
\right]\chi^{c,(n)},
\end{eqnarray}
which suggests that the 0-mode appears
only in $\chi$. 

After integration, the effective four-dimensional
superpotential for the KK state can be written in the form
$W=\sum_{m,n}W^{(m,n)}$, where
\begin{eqnarray}
W^{(m,n)}&=& g^{(m,n)}\Phi \left(X^{(m)}X^{(n)}\right)^2
+ h^{(n)}X^{(n)} Y^2.
\end{eqnarray}
If the matrix ``were'' diagonal ($g^{(m,n)}\simeq g^{(n)}$),
the leading order calculation for the Kaluza-Klein tower ($n\ge 1$)
gives the dissipative coefficient
\begin{eqnarray}
{\cal \gamma}_{KK}&\sim& \sum^{n_{Max}}_{n=1} (g^{(n)})^2 (h^{(n)})^4 
\left[\frac{g^{(n)}\phi}{m^{(n)}_{\chi}}\right]^4
\frac{T^3}{\left[m^{(n)}_{\chi}\right]^2},\nonumber\\ 
\end{eqnarray}
where $n_{Max}\simeq  R M_* $ gives the cut-off at $m_{KK}\sim M_*$.
The off-diagonal elements may contribute to the multiplicity of the diagram.
Let us examine the corrections from the off-diagonal elements.

If one assumes \footnote{This
condition is  not 
mandatory since the extra dimension could be large ($R>M_*^{-1}$).
This possibility will be considered later in this section.}
$g \phi \ll 1/R$,
 the mass eigenstates can be approximated
by the pure KK modes and then one immediately finds
\begin{eqnarray}
{\cal \gamma}_{KK}&\sim& \sum_{n=1}^{n_{Max}}\sum_{m=1}^{n_{Max}} 
\frac{(g^{(m,n)})^6 (h^{(n)})^2(h^{(m)})^2  T^3}{\left(m_{KK}^{(n)}\right)^3
\left(m_{KK}^{(m)}\right)^3}\phi^4\nonumber\\
&\sim& g^6 h^4 T^3 R^6\phi^4
\sum_{n=1}^{n_{Max}} \sum_{n=1}^{n_{Max}} \frac{1}{n^3m^3}\nonumber\\
&\sim& g^6 h^4 T^3 R^6\phi^4\nonumber\\
&<& {\cal \gamma}_0,
\end{eqnarray}
where ${\cal \gamma}_0\equiv C_0 h^4 \frac{T^3}{\phi^2}$ is the dissipative
coefficient for the intermediate 0-mode. 
Therefore, in the above limit ($g \phi \ll 1/R$)
the Kaluza-Klein tower gives negligible
contribution to the dissipative coefficient.
As a result, we find
\begin{eqnarray}
{\cal \gamma}&\simeq& 
C_0 h^4 \frac{T^3}{\phi^2} + C_{KK} g^6 h^4 T^3 R^4\phi^2
\sim  {\cal \gamma}_0,
\end{eqnarray}
where $C_0$ and $C_{KK}$ are numerical coefficients.

Significant enhancement is possible in the opposite
limit, when  $g\phi \gg R^{-1}$.
However, replacing the Gaussian function by $\delta(y)$,
one will find off-diagonal elements of $g^{(m,n)}$, which 
are exactly the same size as the diagonal ones.
For later use, we first estimate the contribution from the diagonal elements.
Here we consider 
\begin{eqnarray}
W^{(n)}&=& g^{(n)}\Phi \left(X^{(n)}\right)^2
+ h^{(n)}X^{(n)} Y^2.
\end{eqnarray}
The leading-order contribution gives   
\begin{eqnarray}
{\cal \gamma}_{KK}&\sim& n_* h^4 \frac{T^3}{\phi^2}\nonumber\\
&\sim& gR h^4 \frac{T^3}{\phi},
\end{eqnarray}
where $n_*\sim gR\phi$ is not the usual cut-off ($n_{Max}$) of the
Kaluza-Klein tower, but obtained from the condition $m_{KK}<g\phi$.
The previous calculation for $g\phi\ll m_{KK}^{(1)}$ suggests that
the Kaluza-Klein states in $n\ge n_*$ is negligible and does not contribute to
the dissipation. 
Finally, we find 
\begin{eqnarray}
{\cal \gamma}&\sim &  C_0 h^4 \frac{T^3}{\phi^2} + \hat{C}_{KK}g R h^4
 \frac{T^3}{\phi}. 
\end{eqnarray}
The above result suggests that there could be an interesting critical point
$\phi_c$, where the dissipation becomes strong due to the Kaluza-Klein
tower.
We can calculate $\phi_c$ as
\begin{equation}
\phi_c \simeq \frac{C_0}{g \hat{C}_{KK}} R^{-1}.
\end{equation}

What is the consequence of $\phi_c$?
If $\phi$ is rolling down on a hilltop potential, it
may feel significant friction when $\phi$ is small\footnote{However,
 $\phi$ must be larger than the temperature of the thermal bath so that the
low-temperature approximation is valid for the calculation.}.
In that case the friction-dominated motion can be prolonged by the
Kaluza-Klein states.
In that way the hilltop inflation~\cite{Hilltop-inf} 
(or the hilltop curvaton~\cite{Hilltop-curv})
could be affected by the Kaluza-Klein tower.
On the other hand, if the field is moving toward the origin, the
Kaluza-Klein enhancement ceases toward $\phi_c$.
This may affect the starting point of the curvaton oscillation and/or 
the Affleck-Dine baryogenesis after inflation~\cite{Warm-curvaton,
Affleck-D, Baryo-matsuda}. 

In the above scenario, the field $\phi$ is localized on the boundary,
which means that the vacuum expectation value
of the field is bounded from above ($\phi\le M_*$).
Therefore, the significant Kaluza-Klein enhancement is possible only
when the extra dimension is large ($RM_*\gg 1$) and $\phi$ is moving in
the region $\phi_c < \phi < M_*$.

Here, our concern is the contributions from
the off-diagonal elements.
To understand the physics related to the off-diagonal elements, let us
consider 2-field model with the superpotential
\begin{eqnarray}
W&=& g\Phi \left(X_1+X_2 \right)^2 +
 h\left(X_1+X_2\right) Y^2.
\end{eqnarray}
Then one can choose $X_\pm\equiv (X_1\pm X_2)/\sqrt{2}$
to obtain
\begin{eqnarray}
W&=& (2g)\Phi X_+^2 + (\sqrt{2}h)X_+ Y^2.
\end{eqnarray}
For the field $X_+$, the dissipation coefficient is multiplied by
$(\sqrt{2}h/h)^4$. 
Applying the same multiplicity, we find
\begin{eqnarray}
\label{KK-min}
{\cal \gamma}_{KK}&\sim& n_*^2 h^4 \frac{T^3}{\phi^2}\nonumber\\
&\sim& (gR)^2 h^4 T^3.
\end{eqnarray}
Note however the weak-coupling expansion is not valid when 
$g_\mathrm{eff}\equiv n_* g \ge 1$ or $h_\mathrm{eff}\equiv
\sqrt{n_*}h\ge 1$.
Therefore, a modest argument would be that Eq.(\ref{KK-min})
certificates significant dissipation, but the 
higher order contributions could be larger than the above result.
At this moment we do not calculate the higher contributions in the
strong limit because it
is not suitable for the purpose of this paper.

\subsection{$\Phi$ and $X$ in the bulk}
Second, assume that $\Phi$ and $X$ are both in the bulk while $Y$ is localized on
the boundary~\cite{bulk-bulk}.
We will introduce the superpotential for the interaction
\begin{eqnarray}
W&=& \left[g \Phi X^2 +h XY^2 \right]\delta(y),
\end{eqnarray}
where the ``point interaction'' is assumed for the Yukawa coupling
$g\Phi X^2$.
This assumption could be removed, however the conventional Yukawa
interaction is forbidden in the ${\cal N}=2$ supersymmetry.  
The effective four-dimensional superpotential that contains the 0-mode 
$\Phi^{(0)}$ and the Kaluza-Klein state is thus given by 
\begin{eqnarray}
W^{(m,n)}&=& g^{(m,n)}\Phi^{(0)} \left(X^{(m)}X^{(n)}\right)^2 +
 h^{(n)}X^{(n)} Y^2.\nonumber\\
\end{eqnarray}
The above superpotential involves the interactions that are required for
the two-step dissipation.
Only the 0-mode of $\phi$ is moving in that model.
All the other fields are supposed to have negligible vacuum
expectation values.

The significant difference appears not in the ``calculation'' but in the
``condition'' that constrains the maximum value of $\phi$.   
In contrast to the previous model, in which $\Phi$ is localized, 
one is not forced to put the condition $\phi \le M_*$.
Therefore, for the diagonal elements we find
${\cal \gamma}_{KK}\sim \hat{C}_{KK} gR h^4 \frac{T^3}{\phi}$.
Again, including the possible multiplicity from the off-diagonal
elements, we find 
${\cal \gamma}_{KK} \sim \check{C}_{KK}(gR)^2 h^4 T^3$, where
$\check{C}_{KK}$ is the numerical coefficient.

\subsection{Dissipation when radius changes}
Note first that $m_{KK}^{(n)}$ is basically the function of $R$, which
could be a  dynamical parameter in the early Universe.
Then a natural question arises: what is the consequence of the
 Kaluza-Klein tower that is changing its spectrum due to the 
dynamical moduli?

In this section we will define the moduli field $\Phi\equiv M_*^2 R$
and consider the motion of $\Phi$.
For simplicity, the canonical kinetic term is assumed for the field
$\Phi$, however this assumption is not mandatory.
$X^c$ and $X$ are in the bulk as usual, and $Y$ is localized on the boundary.
Consider the Sherk-Schwartz (SS) boundary condition~\cite{MSSM-SS}
that replaces $n\rightarrow n+\omega$ in the Fourier expansion of
the bosonic field.
We will introduce the interaction 
\begin{eqnarray}
W&=& h X^c Y^2 \delta(y-\pi R),
\end{eqnarray}
which leads to the auxiliary field equation 
\begin{equation}
F_{X^c}^*=-hY^2\delta(y-\pi R)+\partial_y X.
\end{equation}
The superpotential gives the interaction required for the two-step
dissipation at the leading order.
The crucial interaction is 
\begin{eqnarray}
{\cal L}_\mathrm{int}\sim h^{(n)} m_{KK}^{(n)} \chi^{(n)} y^2,
\end{eqnarray}
where $m_{KK}^{(n)}(\Phi)\simeq  nM_*^2/\Phi$ is the Kaluza-Klein mass
expressed using $\Phi$.
Again, we find the dissipative coefficient
\begin{eqnarray}
{\cal \gamma}_{KK}
&\sim&  \sum_{n=1}^{n_{Max}}
\frac{(m_{KK}^{(n)})^2}{\Phi^2} h^4  
\frac{T^3}{(m_{KK}^{(n)})^2}\nonumber\\
&\sim & h^4 
\frac{T^3}{M_*\Phi},
\end{eqnarray}
where $n_{Max}\sim R M_* \sim \Phi/M_*$ is used in the last line.

Note that the spectrum of the tower is changing
whenever the moduli is time-dependent.
There is no need to consider additional interaction between 
$\Phi$ and $X$.

The above model might be a reminiscence of Casimir force~\cite{Casimir-effect0,
Casimir-matsuda}.
However, in the present model the creation of
the light field causes irreversible process.
They must be discriminated.

\section{Conclusions and discussions}
In this paper we considered two-step dissipation for the
supersymmetric model in which the intermediate field
$X$ is placed in the bulk of the fifth dimension.
In the effective action, the intermediate field has
the tower of Kaluza-Klein states that are coupled to the light fields on
the boundary.

We first considered a simple generalization of 
Ref.\cite{Moss-Xion}.
We first estimated the contribution from the diagonal
interactions.
The Kaluza-Klein tower of the intermediate field $X$ 
enhances the dissipative coefficient when
$\phi>\phi_c\equiv \frac{C_0}{g \hat{C}_{KK}} R^{-1}$.
In that way the dissipative coefficient changes from 
${\cal \gamma}_0= C_0h^4
\frac{T^3}{\phi^2}$ to ${\cal \gamma}\sim
 C_0 h^4\frac{T^3}{\phi^2} + \hat{C}_{KK}g R h^4 \frac{T^3}{\phi}$
for the diagonal elements, and to
${\cal \gamma}\sim C_0 h^4\frac{T^3}{\phi^2} + \check{C}_{KK}(g R)^2 h^4 T^3$ 
when the multiplicity from the off-diagonal elements are included.
More enhancement is expected in the strong ($g_\mathrm{eff},
h_{\mathrm{eff}}>1$) limit. 

Then we considered a model with a moving moduli, in which
the moduli determines the mass of the Kaluza-Klein states.
The result suggests that the back-reaction from the Kaluza-Klein tower
could be crucial for the dynamics of the expanding
extra dimension.
 
We have some comments about the simplification.
In this paper we have disregarded the Kaluza-Klein graviton,
which might be important for the dynamics.
We have also disregarded the decay from 
the higher to the lower Kaluza-Klein states.
Both of those may enhance the dissipative process.

Warped extra dimension has not been considered in this paper.
In that model the fields on the brane can change their mass
when the brane moves along a throat.

\section{Acknowledgment}
T.M thanks N.~Maekawa and S.~Enomoto for many valuable discussions.

\end{document}